\documentstyle[twoside,fleqn,espcrc2,epsf]{article}

\def\spose#1{\hbox to 0pt{#1\hss}}
\def\ltapprox{\mathrel{\spose{\lower 3pt\hbox{$\mathchar"218$}}
 \raise 2.0pt\hbox{$\mathchar"13C$}}}
\def\gtapprox{\mathrel{\spose{\lower 3pt\hbox{$\mathchar"218$}}
 \raise 2.0pt\hbox{$\mathchar"13E$}}}
\def\inapprox{\mathrel{\spose{\lower 3pt\hbox{$\mathchar"218$}}
 \raise 2.0pt\hbox{$\mathchar"232$}}}

\newcommand{\NPB}{Nucl.Phys.B \ }
\newcommand{\PLB}{Phys.Lett.B \ }
\newcommand{\PRD}{Phys.Rev.D \ }

\newcommand{\be}{\begin{equation}}
\newcommand{\ee}{\end{equation}}
\newcommand{\bea}{\begin{eqnarray}}
\newcommand{\eea}{\end{eqnarray}}

\newcommand{\csw}{c_{\rm sw}}

\def\preprints{
\vspace{-10ex}
{\small
\begin{tabbing}
\` Edinburgh 99/13 \\
\` September 1999 \\
\end{tabbing} 
}
\vspace*{0.1in}
}

\title{
\preprints
The Isgur Wise function from the lattice}

\author{
UKQCD collaboration, presented by Gary~Douglas.
\\
\vspace{0.5ex}
\noindent
Department of Physics \&\ Astronomy, University of Edinburgh,
The King's Buildings, Mayfield Road, Edinburgh EH9 3JZ, Scotland}
\begin{document}

\begin{abstract}
We derive the form factors relevant for decays of pseudo-scalar 
mesons corresponding to the semi-leptonic decay 
$\overline{B}\rightarrow Dl\overline{\nu}$. The simulations are performed in the
quenched approximation at $\beta=6.0$ and $\beta=6.2$ using a 
non-perturbatively improved clover action. The slope of the Isgur
Wise function and $|V_{\rm cb}|$ are extracted from the form factors.

\end{abstract}

\maketitle
\section{Introduction}
The calculation of matrix elements corresponding to the semi-leptonic decay of heavy mesons is crucial to the extraction of elements of the CKM matrix. We present results of a lattice study of the decay between heavy-light pseudo-scalar mesons. 
The vector current matrix element can be parametrised in terms of two form factors
\begin{eqnarray}
\lefteqn{\frac{\langle D(v^\prime)|V^\mu|B(v)\rangle}{\sqrt{M_BM_D}}=(v+v^\prime)^\mu h_+(\omega,m_Q,m_{Q^\prime})} \nonumber \\
&&+(v-v^\prime)^\mu h_-(\omega,m_Q,m_{Q^\prime})
\label{eqn:form}
\end{eqnarray}
where $v$ and $v^\prime$ are the meson four velocities and $\omega=v\cdot v^\prime$.
In the heavy quark limit, heavy quark symmetry reduces the two form factors to a single function of the recoil, $\xi(\omega)$, the Isgur Wise function \cite{isgw}. For finite quark mass, there exist two sources of symmetry breaking, one from the exchange of hard gluons allowing resolution of the heavy quark dynamics by the light degrees of freedom, and modifications to the current arising from higher dimensional operators in HQET. The form factors are related to the Isgur Wise function by
\[
h_+(\omega)=\left[1+\beta_+(\omega)+\gamma_+(\omega)\right]\xi(\omega)
\]
\begin{equation}
h_-(\omega)=\left[\beta_-(\omega)+\gamma_-(\omega)\right]\xi(\omega)
\label{eqn:form_sym}
\end{equation}
where $\beta_\pm$ and $\gamma_\pm$ are the radiative and power corrections respectively.
\section{Calculation Details}
The simulation was performed on quenched lattices generated using the Wilson gauge action. The quark propagators were generated from an ${\mathcal O}(a)$ improved Sheikholeslami-Wohlert action with the coefficient $\csw$ determined non-perturbatively \cite{alpha}. The details are summarised in table \ref{tab:sim_dets}.
\vspace{-0.5truecm}
\begin{table}[ht!]
\caption{Simulation details}
\label{tab:sim_dets}
\begin{tabular*}{\columnwidth}{cccccc}\hline
$\beta$ & Volume & $\csw$ & $\kappa_{\rm H}$  & $\kappa_{\rm L}$  \\ \hline \hline
6.0 & $16^3\times 48$ & 1.769 & 0.1123 & 0.13344  \\
& & & 0.1173 & 0.13417   \\
& & & 0.1223 & 0.13455   \\
& & & 0.1273 &   \\ \hline
6.2 & $24^3\times 48$ & 1.614 & 0.1200 & 0.13460   \\
& & & 0.1233 & 0.13510   \\
& & & 0.1266 & 0.13530   \\
& & & 0.1299 &   \\ \hline
\end{tabular*}
\vspace{-0.5truecm}

\end{table}
At $\beta=6.0$ and $\beta=6.2$ we used 305 and 216 configurations respectively. The inverse lattice spacing as determined from the string tension is $a^{-1}=1.89$ GeV for $\beta=6.0$ and $a^{-1}=2.64$ GeV for $\beta=6.2$. The four heavy and three light hopping parameters correpsond to the regions of charm and strange quark masses respectively. The matrix element in equation \ref{eqn:form} is obtained from the large Euclidean time behaviour of the ratio of the three point function over the two meson correlators.
The improvement prescription is completed with the modification of the vector current. Under the non-perturbative scheme, the renormalisation of the vector current is given by
\begin{equation}
V^\mu = Z^{\rm v}_{\rm eff}\left(V^\mu_{\rm latt}+\frac{ac^{\rm v}}{2}[\partial^*_\nu+\partial_\nu]\Sigma^{\mu\nu}\right)
\end{equation}
where $Z^{\rm v}_{\rm eff}=Z^{\rm v}(1+b^{\rm v}am_{\rm q})$ and the improvement coefficients are all known non-perturbatively.

\section{Results for $Z^{\rm v}_{\rm eff}$} 
\begin{figure}[ht!]
\hspace{-3.0truecm}
\vspace{-1.0truecm}
\begin{center}
\leavevmode
\hbox{%
\epsfysize=6.0cm
\epsffile{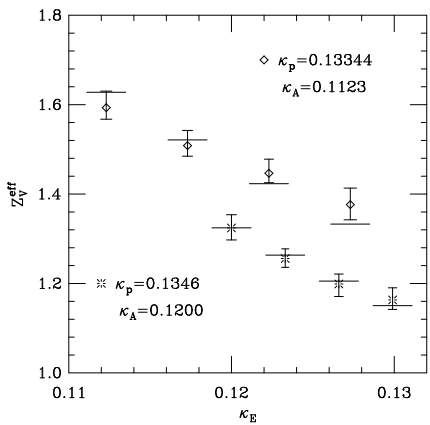}}
\vspace{-1.0truecm}
\caption{$Z^{\rm v}_{\rm eff}$ plotted against the initial heavy quark hopping parameter ($\kappa_{\rm E}$). The horizontal lines represent the non-perturbatively determined values.}
\label{fig:zv_62}
\end{center}
\vspace{-1.0cm}
\end{figure}

At zero recoil, the Isgur Wise function is normalised to one. This condition allows an estimate of the renormalisation constant for the vector current. Hence equation \ref{eqn:form_sym} and the normalisation of $\xi(\omega)$ lead to
\begin{equation}
Z^{\rm v}_{\rm eff}h^{\rm latt}_+(1)=h_+(1)=\left[1+\beta_+(1)+{\mathcal O}\left(\frac{1}{m^2_{\rm Q}}\right)\right]
\end{equation}
where due to Luke's theorem \cite{luke}, the power corrections to $h_+(\omega)$ are suppressed at ${\mathcal O}(\frac{1}{m_{\rm Q}})$. The radiative corrections are perturbative and hence calculable and are obtained using Neubert's prescription \cite{neubert}. The results for $Z^{\rm v}_{\rm eff}$ for a fixed final heavy quark $\kappa_{\rm A}$ and spectator quark $\kappa_{\rm P}$ mass are shown in figure \ref{fig:zv_62}. The agreement with the ALPHA determined values is excellent and thus we conclude that discretisation errors are small.

\section{Heavy Quark Scaling}
\begin{figure}[ht!]
\hspace{-3.0truecm}
\vspace{-1truecm}
\begin{center}
\leavevmode
\hbox{%
\epsfysize=6.5cm
\epsffile{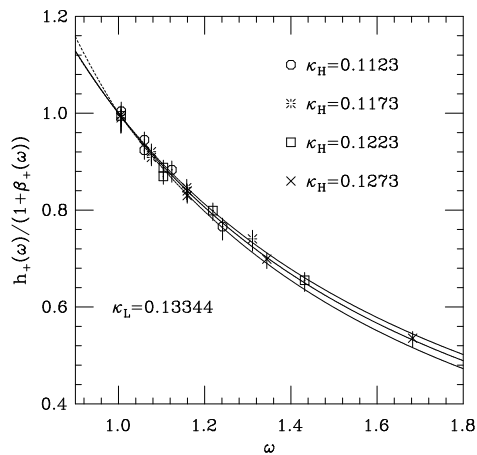}}
\vspace{-1.0truecm}
\caption{The radiatively corrected form factor $h_+(\omega)$ corresponding to degenerate transitions.}
\label{fig:scale}
\end{center}
\vspace{-1.0cm}
\end{figure}

An Isgur-Wise function may be defined as
\begin{equation}
\xi(\omega)=\frac{h_+(\omega)}{1+\beta_+(\omega)}
\end{equation}
To test the heavy quark dependence of the form factors, degenerate transitions are fitted to the Neubert-Rieckert parametrisation \cite{nr} of the Isgur Wise function,
\begin{equation}
\xi(\omega)=\frac{2}{\omega+1}\exp\left[(2\rho^2-1)\frac{1-\omega}{1+\omega}\right]
\end{equation}
The results are shown in figure \ref{fig:scale}. For both datasets, the data lie on the same curve. Independent fitting for each transition shows there is no dependence on heavy quark mass.

\section{Slope of $\xi(\omega)$}
\begin{figure}[ht!]
\hspace{-3.0truecm}
\begin{center}
\leavevmode
\hbox{%
\epsfysize=5.0cm
\epsffile{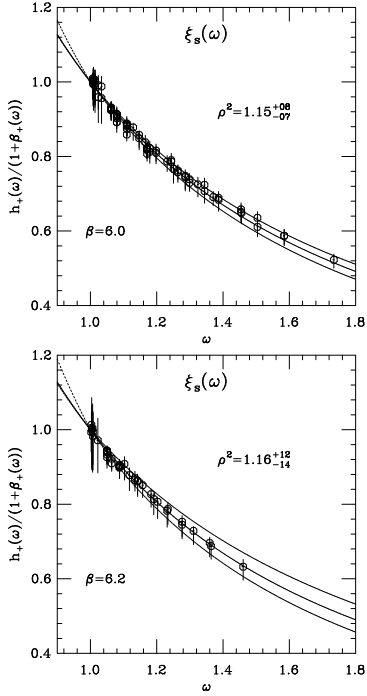}}
\vspace{-1.0truecm}
\caption{Interpolation to the strange quark mass.}
\label{fig:st_60}
\end{center}
\vspace{-1.5cm}
\end{figure}

\begin{figure}[ht!]
\hspace{-3.0truecm}
\begin{center}
\leavevmode
\hbox{%
\epsfysize=5.0cm
\epsffile{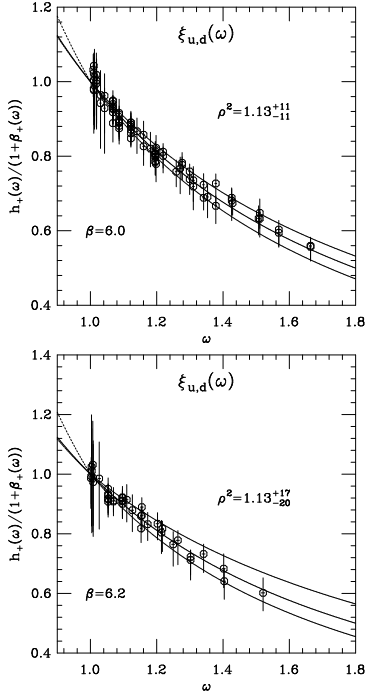}}
\vspace{-1.0truecm}
\caption{Extrapolation to the chiral limit.}
\label{fig:ch_62}
\end{center}
\vspace{-1.5cm}
\end{figure}

The Isgur Wise function corresponding to the physical decays, $\overline{B}_{\rm s}\rightarrow D_{\rm s}l\overline{\nu}$ and $\overline{B}\rightarrow Dl\overline{\nu}$ may be obtained from the extrapolation of the light spectator anti-quark to the strange and chiral limits respectively. With only two values for the spectator, a linear fit is performed in the improved bare quark mass. From an analysis of the meson correlators it is found that at $\beta=6.0$, $\kappa_c=0.13525$ and $\kappa_s=0.13400$, while at $\beta=6.2$, $\kappa_c=0.13583$ and $\kappa_s=0.13493$. The results of the extrapolations are shown in figures \ref{fig:st_60} and \ref{fig:ch_62}. The slopes for each lattice are in excellent agreement and scaling is observed. The slope exhibits a slight dependence on spectator quark mass.
\begin{figure}[ht!]
\hspace{-3.0truecm}
\vspace{-1.0truecm}
\begin{center}
\leavevmode
\hbox{%
\epsfysize=5.0cm
\epsffile{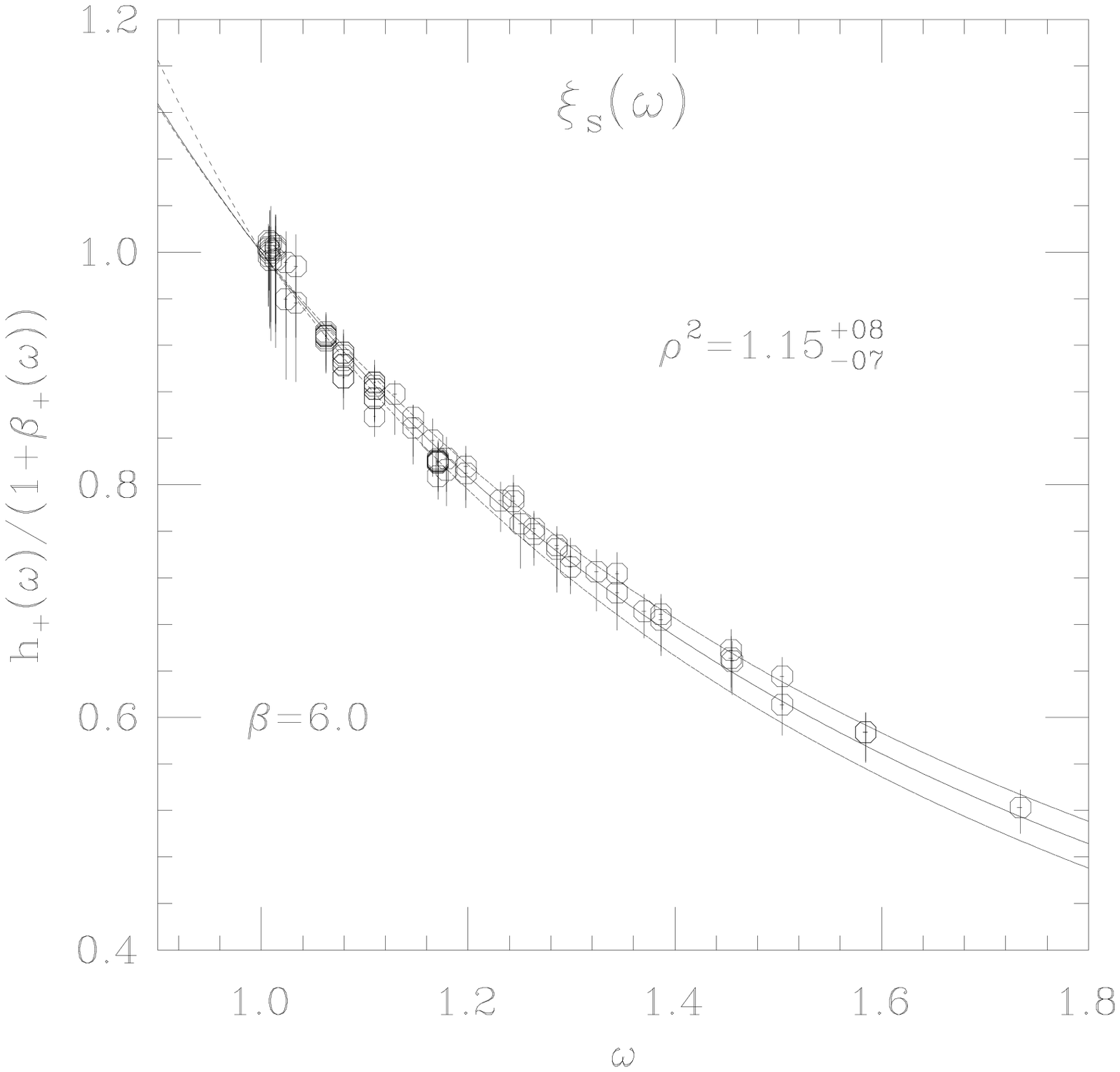}}
\vspace{-2.0truecm}
\end{center}
\end{figure}

\begin{figure}[ht!]
\hspace{-3.0truecm}
\begin{center}
\leavevmode
\hbox{%
\epsfysize=5.0cm
\epsffile{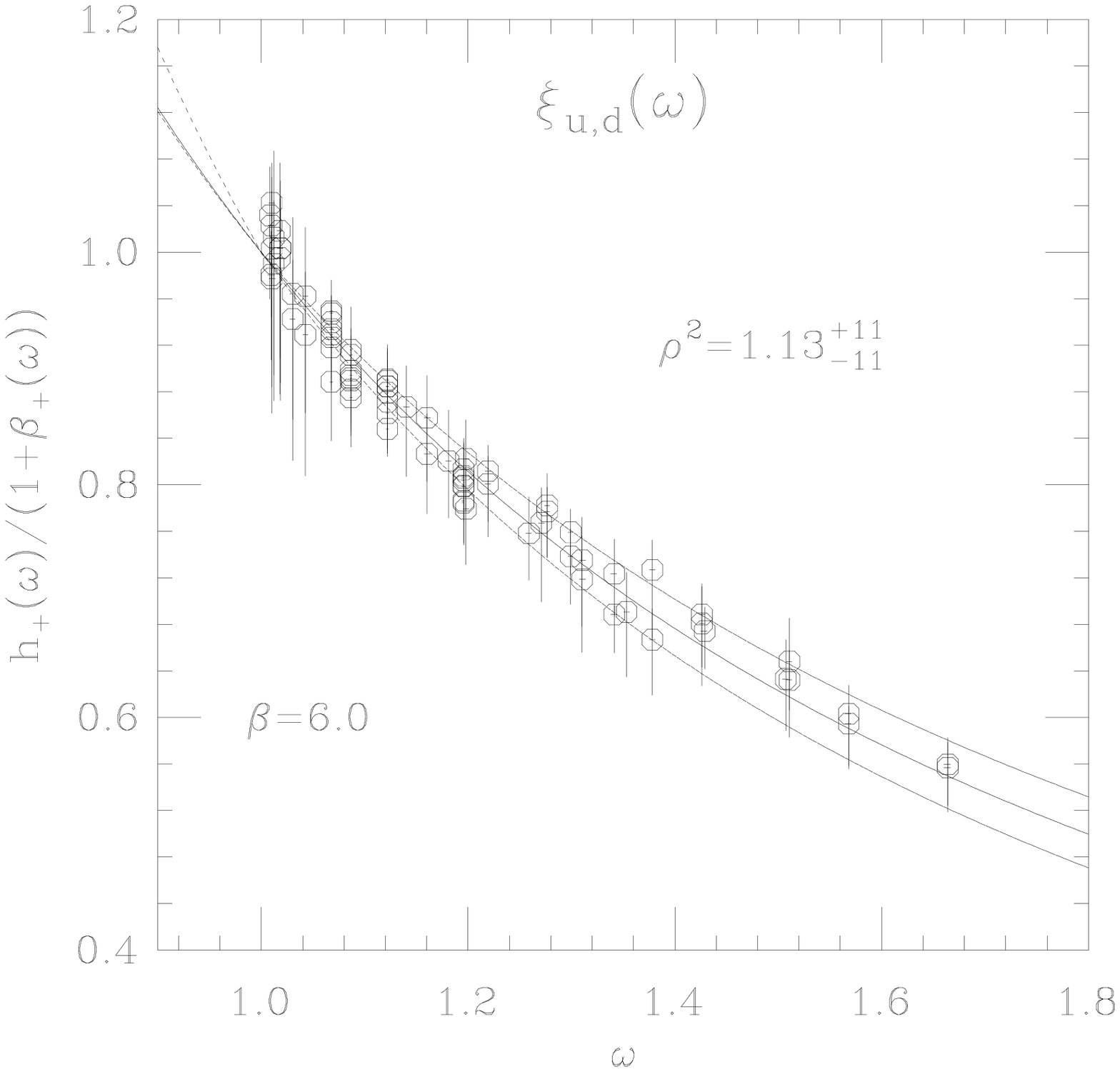}}
\vspace{-1.0truecm}
\end{center}
\vspace{-1.0cm}
\end{figure}

\section{Extracting $|V_{\rm cb}|$}
The CKM element $|V_{\rm cb}|$ may be extracted by comparing the theoretical prediction to the experimentally measured decay rate. However $\overline{B}\rightarrow Dl\overline{\nu}$ is helicity suppressed and hence the measured rate for $\overline{B}\rightarrow D^*l\overline{\nu}$ is used instead. Ignoring kinematic factors, the rate is given by
\begin{eqnarray}
\frac{d\Gamma(\overline{B}\rightarrow D^*l\overline{\nu})}{d\omega}&\propto&\left[1+\beta^{\rm A_1}(1)\right]^2|V_{cb}|^2\nonumber \\
&\times& K(\omega)\xi^2_{\rm u,d}(\omega)
\end{eqnarray}
where $K(\omega)$ is a collection of radiative and power corrections. Making the assumption $K(\omega)=1$  away from the limit of exact heavy quark symmetry, $|V_{\rm cb}|$ is extracted comparing the experimental results from ALEPH \cite{aleph} with our lattice results for $\xi(\omega)$. Hence we find, using Neubert's result for the radiative correction ($\beta^{A_1}(1)=-0.01$),
\[
V_{\rm cb}|_{\beta=6.0}=0.038^{+2+1+2}_{-1-0-2}
\]
\begin{equation}
V_{\rm cb}|_{\beta=6.2}=0.038^{+3+1+2}_{-2-1-2}
\end{equation}
where the first error is statistical, the second is systematic (obtained from analysing the slope of $\xi(\omega)$ for different momentum channels) and the third is experimental. These values are consistent with the world average ($|V_{\rm cb}|=0.0395^{+17}_{-17}$\cite{pdg}).

\section{Acknowledgements}
I would like to acknowledge the support of a PPARC studentship and EPSRC grant GR/K41663.

\end{document}